\documentclass[pdflatex,sn-basic]{sn-jnl}

\usepackage{graphicx}
\graphicspath{{./}}
\usepackage{amsmath,amssymb,amsfonts,mathtools}
\usepackage{physics}
\usepackage{booktabs}
\usepackage{multirow}
\usepackage{array}
\usepackage{xcolor}
\let\orcidlogo\relax
\usepackage{orcidlink}
\usepackage{flafter}
\usepackage{placeins}

\newcommand{\Msun}{M_{\odot}}
\newcommand{\Rsun}{R_{\odot}}
\newcommand{\Me}{M_{\oplus}}
\renewcommand{\Re}{R_{\oplus}}
\newcommand{\Ledd}{L_{\rm Edd}}
\newcommand{\Medd}{\dot{M}_{\rm Edd}}

\newcommand{\rt}{r_{\rm t}}
\newcommand{\rs}{r_{\rm s}}
\newcommand{\rg}{R_{g}}
\newcommand{\MBH}{M_{\rm BH}}

\newcommand{\rp}{r_{p}}

\begin{document}

\articletype{Research Article}

\title[Tidal Disruption of Blanets by Supermassive Black Holes]{Tidal Disruption of Blanets by Supermassive Black Holes: From Test Particles to Planetary-Mass Bodies in Kerr Spacetime}

\author[1]{\fnm{Shreesham} \sur{Pandey}\,\orcidlink{0000-0002-9828-2602}\hspace{0.25em}}
\email{2230257@kmc.du.ac.in}

\author*[1]{\fnm{Sunita} \sur{Singh}}
\email{sunita@kmc.du.ac.in}

\affil*[1]{\orgdiv{Department of Physics}, \orgname{Kirori Mal College, University of Delhi}, \orgaddress{\city{Delhi}, \country{India}}}

\abstract{Planetary-mass bodies formed in situ in active galactic nucleus (AGN) accretion discs, termed ``blanets'', can be driven onto orbits that intersect the tidal-disruption radius of the central supermassive black hole (SMBH). We analyse this process in Kerr spacetime using the standard Mino-time geodesic equations and the Marck (1983) tidal tensor in a parallel-transported tetrad. The blanet is treated as a geodesic test particle, with finite size entering through the tidal-disruption criterion and an estimate of the leading Mathisson--Papapetrou--Dixon correction. For fiducial parameters ($M_p=100\,\Me$, $R_p=6\,\Re$, $\MBH=10^7\,\Msun$), the tidal radius is $\rt=0.82$~AU and $\rt/\rs\approx4.2$, placing disruption at $\rt\approx8.3\,\rg$, close to the innermost stable circular orbit (ISCO). The associated Kerr corrections are therefore non-perturbative. The Hills mass depends on bulk density alone, $M_{\rm Hills}\propto\rho_p^{-1/2}$, and is independent of $M_p$. Rocky compositions ($\rho_p\gtrsim1.4$~g\,cm$^{-3}$) yield a lower Hills mass than a Sun-like star, whereas volatile-rich compositions yield a higher value, making the Hills mass a potential diagnostic of blanet composition. The frozen-in fallback formalism gives $t_{\rm peak}\approx15$~yr and $L_{\rm peak}\approx3.7\times10^{40}$~erg~s$^{-1}$ ($\sim3\times10^{-5}L_{\rm Edd}$), corresponding to a faint, ultraviolet (UV)-peaked transient with a duration of order $10^2$~yr. Scalar resonant relaxation in the surrounding nuclear star cluster may provide a plausible delivery mechanism on $t_{\rm SRR}\sim10^8$--$10^9$~yr, implying a per-AGN event rate of $\dot N_{\rm TDE}\sim10^{-7}$--$10^{-6}\,{\rm yr^{-1}}$. The gravitational-wave strain from sub-Earth-mass debris fragments remains approximately nine orders of magnitude below Laser Interferometer Space Antenna (LISA) sensitivity even at the near-ISCO disruption radius.}


\maketitle

\section{Introduction}\label{sec:intro}

Active galactic nucleus (AGN) accretion discs provide sites of in situ planet formation far from any pre-existing star. Dust grains can grow, settle, and coagulate within the disc midplane to form planetary-mass bodies, termed ``blanets'', on Myr timescales at radii of a few parsecs from the central supermassive black hole (SMBH) \citep{Wada2019,Wada2021}. Once formed, a blanet's long-term evolution is governed by the competition between processes that maintain a stable disc orbit and those that drive inward migration or increase orbital eccentricity. A blanet whose pericentre reaches the SMBH tidal-disruption radius $\rt$ is torn apart in a tidal disruption event (TDE), analogous to a stellar TDE but involving a compact planetary-mass body.

Two features distinguish blanet TDEs from their stellar counterparts. First, blanets are more compact than stars of comparable mass, so $\rt$ is smaller relative to the gravitational radius $\rg\equiv G\MBH/c^2$. For representative parameters, disruption occurs in the strong-field regime at $\rt\sim\text{few}\times\rg$, close to the innermost stable circular orbit (ISCO), where Newtonian tidal theory is insufficient. Second, the ``Hills mass'', defined as the SMBH mass above which $\rt$ lies inside the horizon, depends on the disrupted body's bulk density. Blanet compositions may range from icy and volatile-rich beyond the disc's snow line to rocky and compact, so the Hills mass can exceed the stellar value for volatile-rich bodies and fall below it for rocky bodies.

These features require a fully relativistic treatment. At $\rt\sim\text{few}\times\rg$, the tidal tensor must be evaluated non-perturbatively instead of being expanded in $a_*(\rg/\rt)^{3/2}$. The prograde/retrograde asymmetry is governed by the orbital angular momentum through the Kerr constants of motion. Observational constraints on SMBH spin, including those obtained through relativistic reflection modelling \citep{Reynolds2021,Brenneman2011}, also make spin-dependent predictions relevant. Furthermore, debris circularisation, fallback, and any subsequent gravitational-wave-driven inspiral originate in the strong-field region. We work in the Kerr metric in Boyer--Lindquist coordinates \citep{Kerr1963}, using geometrized units $G=c=1$ unless stated otherwise and restoring physical units in numerical results.

The principal contributions of this study are: (1) a non-perturbative Kerr tidal-tensor treatment of the disruption criterion; (2) the demonstration that the Hills mass depends on blanet bulk density alone and is independent of mass; (3) a self-consistent calculation of the fallback time, luminosity, and duration, identifying blanet TDEs as faint, century-scale UV transients; and (4) an assessment of scalar resonant relaxation as the principal delivery mechanism.

Section~\ref{sec:geodesics} presents the test-particle formalism. Section~\ref{sec:tidal} derives the disruption criterion, tidal radius, density-dependent Hills mass, and non-perturbative Kerr tidal tensor. Section~\ref{sec:stability} addresses orbital stability and the delivery of blanets to $\rt$. Section~\ref{sec:fallback} describes the fallback dynamics and observational signature. Section~\ref{sec:gw} examines the gravitational-wave prospects for post-disruption debris. Section~\ref{sec:rate} estimates the event rate, and Section~\ref{sec:discussion} summarises the results and limitations.

\section{Kerr geodesics and the test-particle framework}\label{sec:geodesics}

\subsection{Metric and constants of motion}

The Kerr metric in Boyer--Lindquist coordinates $(t,r,\theta,\phi)$ is
\begin{align}
  ds^2 = &-\left(1-\frac{2Mr}{\Sigma}\right)dt^2 - \frac{4Mar\sin^2\theta}{\Sigma}\,dt\,d\phi + \frac{\Sigma}{\Delta}dr^2 \nonumber\\
  &+ \Sigma\,d\theta^2 + \left(r^2+a^2+\frac{2Ma^2r\sin^2\theta}{\Sigma}\right)\sin^2\theta\,d\phi^2,
  \label{eq:kerr}
\end{align}
with $\Sigma=r^2+a^2\cos^2\theta$, $\Delta=r^2-2Mr+a^2$, $M$ the SMBH mass and $a=J/M$ its spin parameter ($0\le a_*\equiv a/M\le1$).

A blanet may be treated as a test particle because the mass ratio $q\equiv M_p/\MBH$ is small; for the fiducial parameters below, $q\approx3.00\times10^{-11}$. We therefore normalise the four-velocity to $u^\mu u_\mu=-1$ (i.e.\ $\mu=1$ in the usual notation). The specific energy $E$, axial angular momentum $L$, and Carter constant $Q$ \citep{Carter1968} are defined per unit rest mass. The Carter constant is associated with the hidden symmetry of the Kerr metric and controls the orbit's excursion out of the equatorial plane, with $Q=0$ for purely equatorial motion. The finite mass $M_p$ enters through the mass ratio and the tidal-disruption criterion of Section~\ref{sec:tidal}. Using Mino time $\lambda$, defined by $d\tau=\Sigma\,d\lambda$ \citep{Mino2003}, the geodesic equations separate into
\begin{align}
  \left(\frac{dr}{d\lambda}\right)^2 &= \mathcal{R}(r), \label{eq:mino_r}\\
  \left(\frac{d\theta}{d\lambda}\right)^2 &= \Theta(\theta), \label{eq:mino_th}
\end{align}
with
\begin{align}
  \mathcal{R}(r) &= \bigl[(r^2+a^2)E-aL\bigr]^2 -\Delta\bigl[r^2+(L-aE)^2+Q\bigr], \label{eq:calR}\\
  \Theta(\theta) &= Q-\cos^2\theta\left[a^2(1-E^2)+\frac{L^2}{\sin^2\theta}\right]. \label{eq:calTheta}
\end{align}
No factor of $\Sigma$ appears on the left-hand sides of Eqs.~(\ref{eq:mino_r})--(\ref{eq:mino_th}). Since $dr/d\lambda=\Sigma\,(dr/d\tau)$, substitution into the proper-time equation $\Sigma^2(dr/d\tau)^2=\mathcal{R}(r)$ directly yields Eq.~(\ref{eq:mino_r}). We use $\lambda$ exclusively for the separated radial and polar equations and $\tau$ for physical proper time. Closed-form solutions of Eqs.~(\ref{eq:mino_r})--(\ref{eq:mino_th}) in terms of elliptic functions are given by \citep{FujitaHikida2009}. Equations~(\ref{eq:calR})--(\ref{eq:calTheta}) consistently adopt $\mu=1$, with $E$, $L$, and $Q$ defined per unit rest mass.

\subsection{Finite-size corrections and mass-ratio validity}

Corrections to the point-particle geodesic from the blanet's finite size and any internal angular momentum are governed by the Mathisson--Papapetrou--Dixon (MPD) equations \citep{Papapetrou1951},
\begin{equation}
  \frac{Du^\mu}{d\tau} = -\frac{1}{2}R^\mu{}_{\nu\alpha\beta}u^\nu S^{\alpha\beta},
  \label{eq:MPD}
\end{equation}
where $S^{\alpha\beta}$ is the spin tensor. For a non-rotating, structurally rigid blanet, the leading spin-curvature correction relative to geodesic motion scales as $|S|/(M_p r)\times(M_p/M)\sim(R_p/\rt)^2\,q$. This estimate follows from the bound $|S|\lesssim M_p R_p^2\Omega_{\rm orb}$ and the local curvature length scale. The correction is $\approx2.9\times10^{-18}$ for the adopted parameters and is neglected. Rapid rotation or strong tidal deformation near disruption would require an extended-body treatment.

\subsection{Innermost stable circular orbit and apsidal precession}

The equatorial ISCO radius follows the standard Bardeen--Press--Teukolsky form \citep{Bardeen1972},
\begin{equation}
  r_{\rm ISCO} = M\Bigl\{3+Z_2 \mp \bigl[(3-Z_1)(3+Z_1+2Z_2)\bigr]^{1/2}\Bigr\},
  \label{eq:risco}
\end{equation}
with $Z_1=1+(1-a_*^2)^{1/3}\bigl[(1+a_*)^{1/3}+(1-a_*)^{1/3}\bigr]$ and $Z_2=(3a_*^2+Z_1^2)^{1/2}$. The upper and lower signs correspond to retrograde and prograde orbits, respectively. The ISCO ranges from $r_{\rm ISCO}=M$ for a maximally spinning prograde orbit to $9M$ for a maximally spinning retrograde orbit, with $r_{\rm ISCO}=6M$ at $a_*=0$. Equation~(\ref{eq:risco}) provides a comparison scale for the tidal radius in Section~\ref{subsec:kerr_tidal}. For a near-circular orbit at radius $r_0\gg M$, the leading relativistic apsidal precession per orbit is
\begin{equation}
  \Delta\phi_{\rm prec} \approx \frac{6\pi M}{r_0},
  \label{eq:precession}
\end{equation}
which is applicable to the parsec-scale weak-field region considered in Section~\ref{sec:stability}, where $r_0/M\sim10^6$--$10^7$. Near $\rt$, the non-perturbative treatment of Section~\ref{subsec:kerr_tidal} is used instead.

\section{Tidal disruption in Kerr spacetime}\label{sec:tidal}

\subsection{Disruption criterion}

A blanet is disrupted when the tidal acceleration across it exceeds its self-gravity. In a local orthonormal frame comoving with the blanet, the tidal tensor $C_{\hat\imath\hat\jmath}\equiv R_{\hat\imath\hat 0\hat\jmath\hat 0}$ is the electric part of the Weyl tensor projected onto that frame (the magnetic part does not couple to geodesic deviation for a freely-falling frame at this order) and satisfies the disruption condition
\begin{equation}
  |C_{\hat r\hat r}|\,R_p \gtrsim \frac{GM_p}{R_p^2},
  \label{eq:disruption_cond}
\end{equation}
i.e.\ the radial stretching term dominates the blanet's own surface gravity. We compare only the radial eigenvalue against self-gravity because $\Lambda_1>0$ is the unique \emph{stretching} (as opposed to compressive) eigenvalue for these orbits (Eqs.~\ref{eq:Lambda1}--\ref{eq:Lambda3} below): the transverse eigenvalues $\Lambda_2,\Lambda_3<0$ compress the body and do not drive disruption, so the radial direction is always the first to fail. For a purely radial (zero angular momentum) plunge in Schwarzschild spacetime this reduces to the standard result
\begin{equation}
  C_{\hat r\hat r} = -\frac{2M}{r^3}, \qquad C_{\hat\theta\hat\theta}=C_{\hat\phi\hat\phi}=+\frac{M}{r^3},
  \label{eq:tidal_schw}
\end{equation}
which is exact at all $r$ for this orbit (not merely a weak-field limit).

\subsection{The Kerr tidal tensor: a non-perturbative treatment}\label{subsec:kerr_tidal}

Specializing the parallel-transported Marck tidal tensor to the equatorial timelike geodesics considered here and diagonalizing the resulting spatial tidal matrix, following the procedure of \citep{Marck1983} and \citep{Kesden2012}, gives the eigenvalues
\begin{align}
  \Lambda_1 &= \frac{2M}{r^3}\left(1+\frac{3K}{\tilde r^{2}}\right), \label{eq:Lambda1}\\
  \Lambda_2 &= -\frac{M}{r^3}\left(1+\frac{3K}{\tilde r^{2}}\right), \label{eq:Lambda2}\\
  \Lambda_3 &= -\frac{M}{r^3}, \label{eq:Lambda3}
\end{align}
where $\tilde r\equiv r/M$ and $K\equiv(L-a_*E)^2+Q$ (in units of $M^2$) is the combination of Kerr constants of motion that enters the radial potential $\mathcal{R}(r)$. The quantity $K$ is distinct from the Carter constant $Q$ and reduces to $Q$ only when $L=a_*E$. Appendix~\ref{app:derivation} summarises the tidal-tensor construction and its limiting cases. A complementary ZAMO-frame formulation for stationary axisymmetric spacetimes is presented by \citep{XinMummery2026}. The Marck tetrad is used here because it is comoving and remains valid along the eccentric pericentre passage relevant to disruption. The prograde/retrograde asymmetry enters through the sign of $L$ in $K$. Equations~(\ref{eq:Lambda1})--(\ref{eq:Lambda3}) reduce to Eq.~(\ref{eq:tidal_schw}) as $K\to0$ at any $r$, and to the standard Newtonian/Schwarzschild weak-field form as $\tilde r\to\infty$.

\begin{figure}[htbp]
  \centering
  \includegraphics[width=\columnwidth]{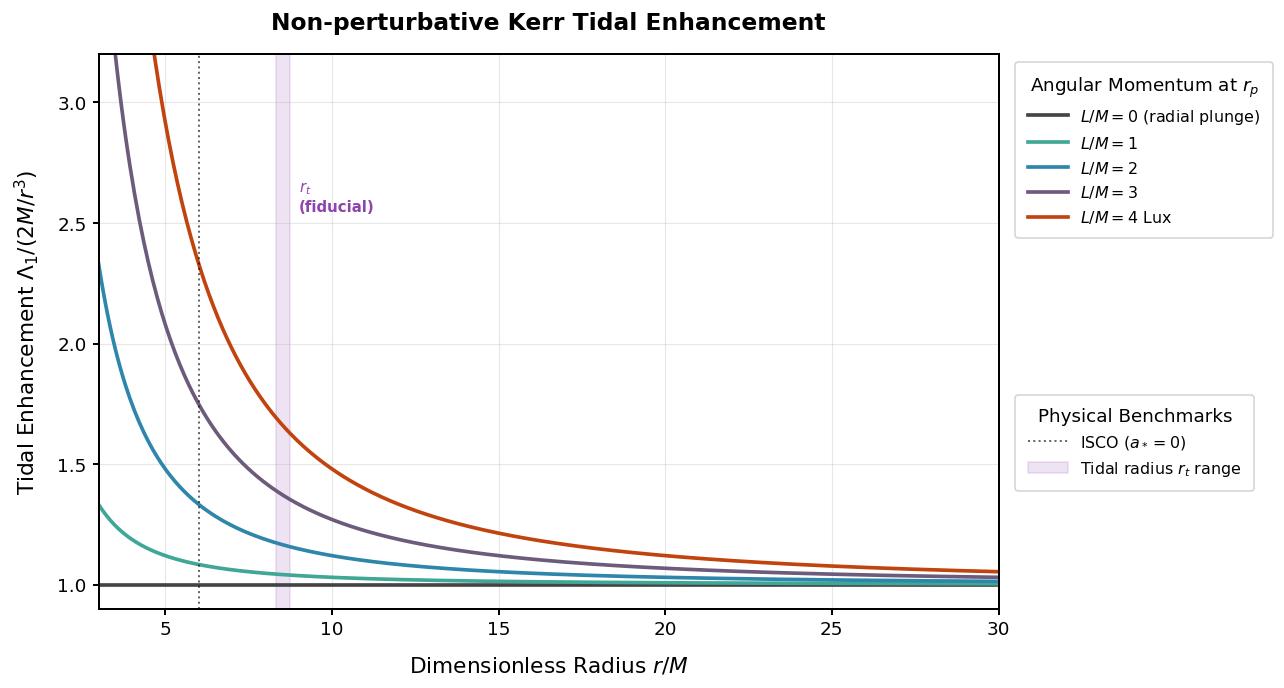}
  \caption{Non-perturbative Kerr enhancement of the radial tidal eigenvalue, $\Lambda_1/(2M/r^3)=1+3K/\tilde r^2$ (Eq.~\ref{eq:Lambda1}), as a function of radius for representative pericentre angular momenta. At $\rt$ (purple band), the enhancement reaches $\sim70\%$ for the plotted $L/M=4$ case, which represents a high-angular-momentum orbit within the adopted illustrative bracket rather than a physically preferred value.}
  \label{fig:tidal_enhancement}
\end{figure}

Equations~(\ref{eq:Lambda1})--(\ref{eq:Lambda3}) are evaluated at the blanet's pericentre $\rp$, rather than at its parsec-scale formation radius, in a tetrad parallel-transported along the generally eccentric disruption orbit. Because $K$ and $\tilde r$ depend on the specific orbit, the self-consistent tidal radius satisfies
\begin{equation}
  r_t^{\rm GR} = R_p\left[\left(1+\frac{3K}{\tilde r^{2}}\right)\frac{2\MBH}{M_p}\right]^{1/3}, \qquad \tilde r = r_t^{\rm GR}/M,
  \label{eq:rt_GR_selfconsistent}
\end{equation}
which is solved iteratively. The representative range $L/M\in[0,4]$ with $Q=0$ spans a radial plunge through bound orbits near the ISCO; it is not intended to represent a population distribution. A population-level calculation would require the distribution of $(L,Q)$ generated by the resonant-relaxation channel of Section~\ref{sec:stability}. Table~\ref{tab:enhancement} gives $r_t^{\rm GR}/r_t^{\rm Newtonian}$ across this range. For the representative case $L/M\approx2$, $Q=0$, and $K\approx4$, Eq.~(\ref{eq:rt_GR_selfconsistent}) gives $r_t^{\rm GR}/M=8.74$, approximately $5\%$ above the Newtonian value in Eq.~(\ref{eq:rt_physical}). At fixed $r$, the radial tidal eigenvalue $\Lambda_1$ is enhanced by as much as $70\%$ over $L/M\in[0,4]$, as shown in Figure~\ref{fig:tidal_enhancement}. Equation~(\ref{eq:rt_GR_selfconsistent}) defines the disruption radius used below. A population-specific calculation would additionally resolve the prograde and retrograde distributions of $L$ and $Q$ for each SMBH spin.

\begin{table}[htbp]
\centering
\caption{Self-consistent GR enhancement of the tidal radius (Eq.~\ref{eq:rt_GR_selfconsistent}) over the representative bracket $L/M\in[0,4]$, $Q=0$.}
\label{tab:enhancement}
\begin{tabular}{@{}ccc@{}}
\toprule
$L/M$ & $r_t^{\rm GR}/M$ & $r_t^{\rm GR}/r_t^{\rm Newtonian}$ \\
\midrule
0 & 8.33 & 1.000 \\
1 & 8.45 & 1.014 \\
2 & 8.75 & 1.050 \\
3 & 9.14 & 1.098 \\
4 & 9.58 & 1.150 \\
\bottomrule
\end{tabular}
\end{table}

\begin{figure}[htbp]
  \centering
  \includegraphics[width=\columnwidth]{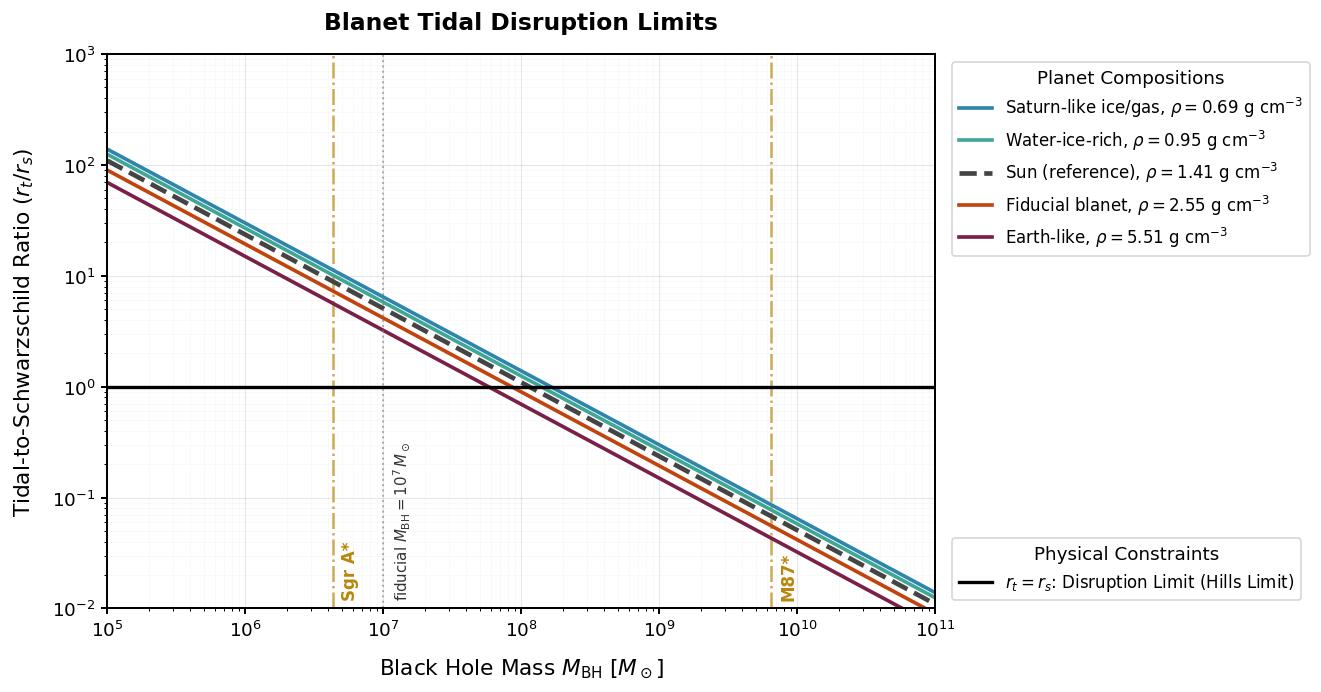}
  \caption{Tidal-to-Schwarzschild radius ratio $\rt/\rs$ as a function of $\MBH$ for representative blanet bulk densities (Eq.~\ref{eq:hills_blanet}). The horizontal line marks the Hills limit $\rt=\rs$, above which disruption is impossible. Vertical lines identify Sgr~A* ($4.3\times10^6\,\Msun$) and M87* ($6.5\times10^9\,\Msun$), while the fiducial $10^7\,\Msun$ SMBH is also indicated. Volatile-rich compositions permit disruption by more massive black holes than rocky compositions.}
  \label{fig:hills_composition}
\end{figure}

For context, Figure~\ref{fig:hills_composition} presents the density-dependent tidal-disruption limits derived in Section~\ref{subsec:hills}, while Figure~\ref{fig:tidal_enhancement} isolates the strong-field Kerr enhancement discussed above.

Figure~\ref{fig:tidal_enhancement} also shows that the relativistic correction is concentrated near the black hole and depends strongly on the orbital constants. Across the fiducial disruption band, the plotted enhancement ranges from unity for a radial plunge to approximately $1.7$ for the representative high-angular-momentum case $L/M=4$, while the representative $L/M=2$ curve gives an enhancement of about $17\%$. These values sample the adopted orbital bracket and do not identify either angular momentum as physically special. All curves approach the Schwarzschild/Newtonian value at large radius. Thus, the modest change in the self-consistent tidal radius listed in Table~\ref{tab:enhancement} can coexist with a substantially larger local enhancement of the stretching eigenvalue.

\FloatBarrier

\subsection{Newtonian tidal radius and the density-only Hills mass}\label{subsec:hills}

At leading (Newtonian) order, $K\to0$ and Eq.~(\ref{eq:rt_GR_selfconsistent}) reduces to the standard tidal-disruption radius \citep{Hills1975,Rees1988},
\begin{equation}
  \rt = R_p\left(\frac{\MBH}{M_p}\right)^{1/3},
  \label{eq:rt_physical}
\end{equation}
For the fiducial blanet parameters used throughout this paper, $M_p=100\,\Me$, $R_p=6\,\Re$ (bulk density $\rho_p=2.55\,{\rm g\,cm^{-3}}$), and $\MBH=10^7\,\Msun$, Eq.~(\ref{eq:rt_physical}) gives
\begin{equation}
  \rt = 1.23\times10^{13}\,{\rm cm} = 0.82\,{\rm AU}, \qquad \rt/\rs = 4.16,
  \label{eq:rt_numeric}
\end{equation}
with $\rs=2G\MBH/c^2$. Because $\rt/\rs$ is of order a few, the disruption pericentre lies at $\rt\approx8.3\,\rg$, comparable to the ISCO (Eq.~\ref{eq:risco}) and within the regime requiring the non-perturbative treatment of Section~\ref{subsec:kerr_tidal}.

The SMBH mass above which $\rt$ recedes inside $\rs$ altogether (the Hills mass) follows by setting $\rt=\rs$ in Eq.~(\ref{eq:rt_physical}) and using $R_p=(3M_p/4\pi\rho_p)^{1/3}$:
\begin{equation}
  M_{\rm Hills} = \left(\frac{R_p c^2}{2G}\right)^{3/2} M_p^{-1/2} = \left(\frac{c^2}{2G}\right)^{3/2}\left(\frac{3}{4\pi\rho_p}\right)^{1/2}.
  \label{eq:hills_blanet}
\end{equation}
The dependence on $M_p$ cancels exactly, so the Hills mass of a self-gravitating body depends only on its bulk density $\rho_p$ under this criterion. Table~\ref{tab:hills_composition} evaluates Eq.~(\ref{eq:hills_blanet}) across representative blanet densities and includes a Sun-like star ($\rho_\odot=1.41\,{\rm g\,cm^{-3}}$) for comparison. The resulting stellar value agrees with the standard Hills mass of $\sim10^8\,\Msun$ \citep{Hills1975}.

\begin{table}[htbp]
\centering
\caption{Hills mass as a function of bulk density (Eq.~\ref{eq:hills_blanet}), independent of $M_p$. The Sun is included as a stellar reference.}
\label{tab:hills_composition}
\begin{tabular}{@{}lcc@{}}
\toprule
Composition & $\rho_p$ (g\,cm$^{-3}$) & $M_{\rm Hills}$ ($\Msun$) \\
\midrule
Saturn-like ice/gas       & 0.69 & $1.63\times10^8$ \\
Water-ice-rich            & 0.95 & $1.39\times10^8$ \\
\textbf{Sun (reference)}  & \textbf{1.41} & $\mathbf{1.14\times10^8}$ \\
Neptune-like              & 1.64 & $1.06\times10^8$ \\
\textbf{Fiducial blanet}    & \textbf{2.55} & $\mathbf{8.50\times10^7}$ \\
Earth-like                & 5.51 & $5.78\times10^7$ \\
\bottomrule
\end{tabular}
\end{table}

Blanet Hills masses therefore straddle the stellar value. Volatile-rich compositions with $\rho_p\lesssim1.4\,{\rm g\,cm^{-3}}$, as may arise through ice accretion beyond the disc's snow line \citep{Wada2021}, yield higher Hills masses, whereas rocky compositions yield lower values. This density dependence may provide a diagnostic of blanet composition. Mapping it onto blanet mass for a specific formation scenario requires a mass--radius relation, such as the broken power law of \citep{ChenKipping2017}, across the relevant $20$--$3000\,\Me$ range. The resulting tidal-to-Schwarzschild radius ratios across representative bulk densities are shown in Figure~\ref{fig:hills_composition}.

\FloatBarrier

\subsection{Formation radius vs.\ disruption pericentre}

The tidal radius $\rt$ in Eq.~(\ref{eq:rt_numeric}) denotes the disruption pericentre and is distinct from the parsec-scale formation radius discussed in Section~\ref{sec:stability}. Relativistic effects are evaluated at $\rp\simeq\rt$, with $R_p$ denoting the blanet radius and $\rp$ the orbital pericentre.

\section{Orbital stability and the destabilisation channel}\label{sec:stability}

\subsection{The safe zone}

Blanets can form and persist where they are neither disrupted nor ejected, $\rt<r_{\rm safe}<r_{\rm disk}$, with $r_{\rm disk}\sim10$--$100$~pc representing the outer disc edge. Formation models place blanets primarily within $\sim1$--$20$~pc \citep{Wada2019,Wada2021}. The precise range depends on the adopted surface-density, temperature, and viscosity profiles, so the values above are treated as fiducial. At these radii, the Keplerian period is $P_{\rm orb}\sim3\times10^5\,{\rm yr}\,(a/5\,{\rm pc})^{3/2}(\MBH/10^7\Msun)^{-1/2}$, as used in Eq.~(\ref{eq:t_disc_prec}).

\subsection{Lense--Thirring precession}

Frame dragging induces nodal precession at rate
\begin{equation}
  \Omega_{\rm LT} = \frac{2Ma}{r^3},
  \label{eq:OmLT}
\end{equation}
with precession timescale $t_{\rm LT}=2\pi/\Omega_{\rm LT}$. At the safe-zone scale this is
\begin{equation}
  t_{\rm LT} \approx 6.2\times10^{15}\,{\rm yr}\left(\frac{r}{5\,{\rm pc}}\right)^3\left(\frac{\MBH}{10^7\Msun}\right)^{-1}\left(\frac{a_*}{0.9}\right)^{-1},
  \label{eq:tLT}
\end{equation}
for $r=5$~pc, $\MBH=10^7\,\Msun$, and $a_*=0.9$. Lense--Thirring precession is therefore too slow to affect the parsec-scale dynamics.

\subsection{Type-I migration}

Gas-disc torques drive inward migration on the standard Type-I timescale \citep{Tanaka2002},
\begin{equation}
  t_{\rm migI} = \frac{C_{\rm mig}}{\Omega_K}\,\frac{\MBH^2}{M_p M_d}\left(\frac{H}{r}\right)^2, \qquad M_d=\pi\Sigma r^2,
  \label{eq:t_mig}
\end{equation}
with $C_{\rm mig}\approx0.8$ in three dimensions. Because $M_d$ depends on the AGN disc surface density $\Sigma(r)$, the migration timescale is model-dependent. Equation~(\ref{eq:t_mig}) shows that $t_{\rm migI}\lesssim\tau_{\rm AGN}=10^8$~yr at $a_0=5$~pc requires $\Sigma\gtrsim10^5$--$10^6\,{\rm g\,cm^{-2}}$ for aspect ratios $H/r=0.02$--$0.10$ (Table~\ref{tab:migration}). Determining whether a marginally self-gravitating AGN disc \citep{SirkoGoodman2003,Thompson2005} reaches these surface densities requires a specified disc-structure model, for example one implemented with \texttt{pAGN} \citep{Gangardt2024}.

\begin{table}[htbp]
\centering
\caption{Surface density $\Sigma$ required for $t_{\rm migI}=10^8$~yr at $a_0=5$~pc (Eq.~\ref{eq:t_mig}), for the fiducial blanet.}
\label{tab:migration}
\begin{tabular}{@{}lc@{}}
\toprule
$H/r$ & $\Sigma_{\rm required}$ (g\,cm$^{-2}$) \\
\midrule
0.02 & $1.5\times10^5$ \\
0.05 & $9.3\times10^5$ \\
0.10 & $3.7\times10^6$ \\
\bottomrule
\end{tabular}
\end{table}

\FloatBarrier

\subsection{Eccentricity-pumping channels and their competition with relativistic and disc precession}\label{subsec:destabilisation}

For a blanet to reach $\rt$ from a near-circular safe-zone orbit, its eccentricity must grow until the pericentre distance falls to $\rt$. We examine three candidate channels and the precession mechanisms that compete against each.

\emph{Relativistic apsidal precession} (Eq.~\ref{eq:precession}) detunes secular eccentricity-pumping resonances when its timescale becomes shorter than the pumping timescale. At $a_0=5$~pc, $t_{\rm GR}\approx1.15\times10^{12}$~yr at $e\approx0$ and decreases to $\sim2.3\times10^{10}$~yr at $e=0.99$. It therefore becomes important only after the eccentricity has grown substantially.

\emph{Disc-induced precession}, from the disc's own enclosed mass, scales as (\citep{Merritt2013}; consistent with the independent scaling found in galactic-centre gas-disc precession studies)
\begin{equation}
  T_{\rm disc} \sim P_{\rm orb}\,\frac{\MBH}{M_d(<a)}.
  \label{eq:t_disc_prec}
\end{equation}
At $a_0=5$~pc ($P_{\rm orb}=3.3\times10^5$~yr), this ranges from $3.3\times10^9$~yr ($M_d=10^3\,\Msun$) to $3.3\times10^6$~yr ($M_d=10^6\,\Msun$): competitive with $\tau_{\rm AGN}$ only for a substantial enclosed disc mass, again requiring the explicit $\Sigma(r)$ noted above.

\emph{Kozai--Lidov (KL) oscillations} \citep{Kozai1962,Lidov1962} from a single inclined perturber at $r_1\gg a_0$ operate on
\begin{equation}
  T_{\rm KL} = \frac{2\pi\sqrt{GM_{\rm BH}}}{Gm_1}\,\frac{r_1^3}{a_0^{3/2}}(1-e_1^2)^{3/2}.
  \label{eq:t_KL}
\end{equation}
At $r_1=5$--$10$~pc, a KL timescale of $10^7$--$10^8$~yr requires $m_1\sim3\times10^4$--$3\times10^6\,\Msun$, corresponding to an intermediate-mass black hole (IMBH). By contrast, a $10$--$100\,\Msun$ perturber gives $T_{\rm KL}\gtrsim10^{11}$~yr, which exceeds $\tau_{\rm AGN}$ and is dominated by relativistic apsidal precession. Single-perturber KL oscillations are therefore relevant only in systems that host a sufficiently massive companion.

\emph{Scalar resonant relaxation} \citep{RauchTremaine1996}, driven by the ambient nuclear star cluster, operates on $t_{\rm SRR}\sim10^8$--$10^9$~yr in representative galactic-nucleus models \citep{BarOrFouvry2018}. This range is competitive with $\tau_{\rm AGN}=10^8$~yr and does not require a single massive perturber. Its efficiency depends on the nuclear star cluster density profile and orbital coherence time, so the quoted range represents an order-of-magnitude estimate rather than a system-specific prediction. Scalar resonant relaxation is adopted as the primary delivery mechanism in Section~\ref{sec:rate}, while Type-I migration (Eq.~\ref{eq:t_mig}) and KL cycling (Eq.~\ref{eq:t_KL}) remain possible secondary channels in suitable systems.

Together, these results define a coherent evolutionary pathway. Dust growth and coagulation in the AGN disc produce blanets at radii of a few parsecs, after which they may occupy long-lived, approximately circular orbits within the $\sim1$--$20$~pc safe zone. At these radii, Lense--Thirring precession is negligible, while the importance of Type-I migration depends on the local disc structure. Over longer timescales, scalar resonant relaxation may provide the most generally applicable mechanism for increasing orbital eccentricity, with $t_{\rm SRR}\sim10^8$--$10^9$~yr. When the pericentre reaches $r_p\simeq\rt\approx8.3\,\rg$, the blanet undergoes strong-field tidal disruption close to the ISCO. The subsequent debris fallback produces the faint, ultraviolet-peaked transient discussed in Section~\ref{sec:fallback}, with an observable duration of order $10^2$~yr.

\begin{figure}[htbp]
  \centering
  \includegraphics[width=\columnwidth]{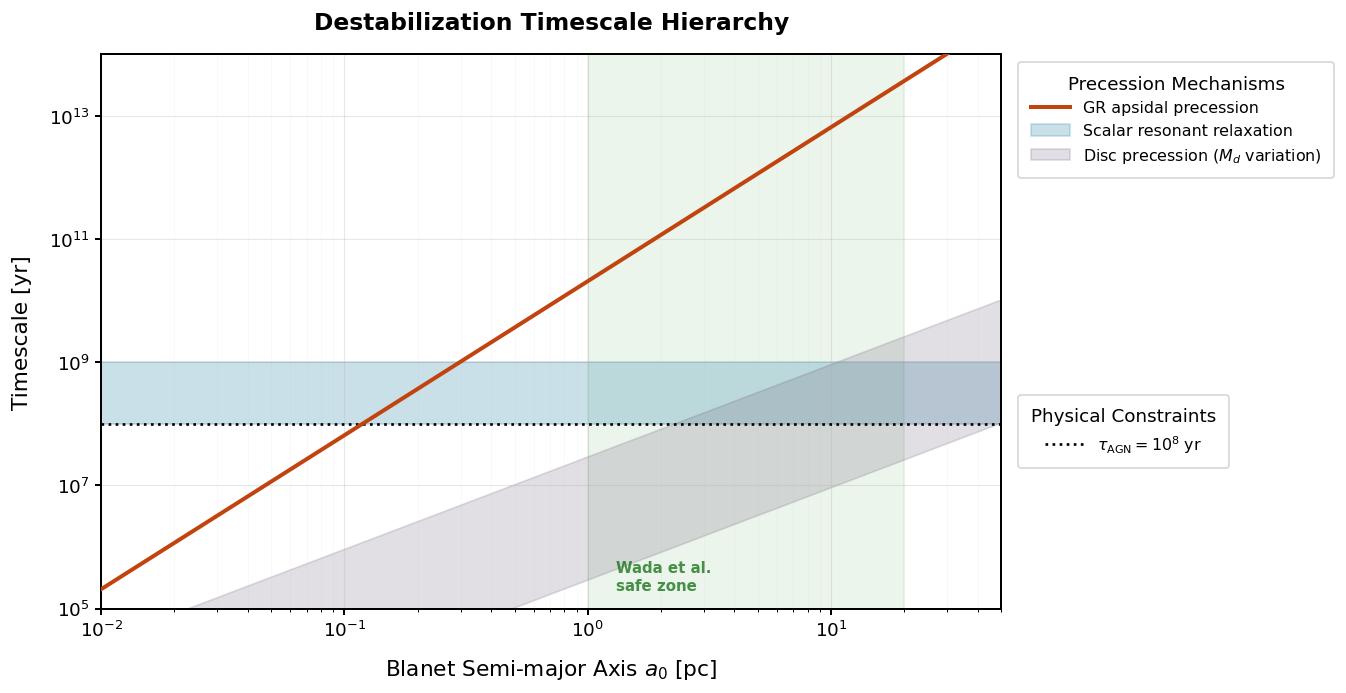}
  \caption{Destabilisation timescale hierarchy as a function of blanet semi-major axis. GR apsidal precession (Eq.~\ref{eq:precession}) is negligible at safe-zone scales. The disc-precession band (Eq.~\ref{eq:t_disc_prec}) spans a representative range of enclosed disc masses, while scalar resonant relaxation is competitive with $\tau_{\rm AGN}=10^8$~yr (dotted line). The shaded green region marks the approximate blanet safe zone.}
  \label{fig:hierarchy}
\end{figure}

Figure~\ref{fig:hierarchy} summarises the timescale hierarchy underlying the middle step, as a function of blanet semi-major axis.

\subsection{The ISCO across spin, and where disruption actually sits}

\begin{figure}[htbp]
  \centering
  \includegraphics[width=\columnwidth]{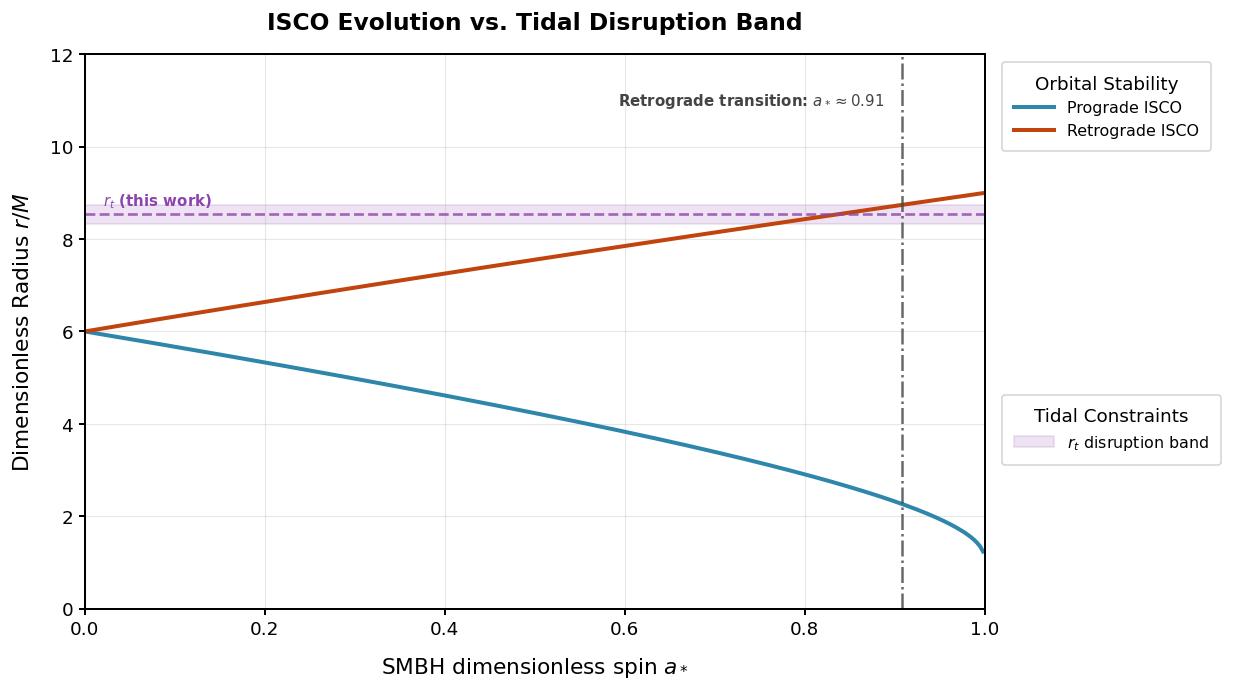}
  \caption{Prograde and retrograde ISCO radii as functions of SMBH spin $a_*$, with the disruption radius $\rt\approx8.3$--$8.7\,M$ overlaid. The retrograde ISCO enters the disruption band near $a_*\approx0.75$ and reaches its upper edge near $a_*\approx0.91$; the ``retrograde transition'' marker in the figure denotes this high-spin boundary.}
  \label{fig:eff_potential}
\end{figure}

Because $\rt\approx8.3$--$8.7\,M$ (Section~\ref{subsec:kerr_tidal}) is close to the ISCO, its position relative to the ISCO depends on spin and orbital orientation. Figure~\ref{fig:eff_potential} shows the prograde and retrograde ISCO radii (Eq.~\ref{eq:risco}) across the full spin range, together with the $\rt$ band. For low-to-moderate spin, $\rt$ lies outside the ISCO for both orientations. The retrograde ISCO reaches the $\rt$ band near $a_*\approx0.75$; at higher spins, a retrograde encounter may plunge directly rather than produce a resolved tidal disruption. Quantitative disruption rates at high spin therefore require separate prograde and retrograde orbital distributions.

\FloatBarrier

\section{Fallback dynamics and observational signature}\label{sec:fallback}

\subsection{Peak fallback time}

In the frozen-in approximation, the spread in specific orbital energy across the disrupted body at pericentre sets the fallback timescale \citep{Lacy1982,Rees1988}:
\begin{equation}
  t_{\rm peak} = \frac{2\pi}{\sqrt{G\MBH}}\left(\frac{\rt^2}{2R_p}\right)^{3/2} \approx 0.11\,{\rm yr}\left(\frac{\MBH}{10^6\Msun}\right)^{1/2}\left(\frac{M_p}{\Msun}\right)^{-1}\left(\frac{R_p}{\Rsun}\right)^{3/2}.
  \label{eq:t_peak_general}
\end{equation}
For the fiducial blanet parameters at $\MBH=10^7\,\Msun$,
\begin{equation}
  t_{\rm peak} \approx 15.2\,{\rm yr},
  \label{eq:t_peak_blanet}
\end{equation}
Since $t_{\rm peak}\propto\MBH^{1/2}M_p^{-1}R_p^{3/2}$, the value $15.2$~yr applies specifically to the adopted fiducial system; other blanet masses, radii, and SMBH masses rescale according to Eq.~(\ref{eq:t_peak_general}). It may be compared with $\sim129.5$~days for a Sun-like star disrupted by an SMBH of the same mass (Table~\ref{tab:comparison}).

\subsection{Peak fallback rate and luminosity}

The fallback rate $\dot M(t)$ computed here is the ballistic mass-return rate, not the emergent luminosity: the two are connected only through the assumed constant radiative efficiency $\eta=0.1$, and in reality $\eta$ can vary during circularisation, particularly for the sub-Eddington, extended circularisation expected here. With that caveat, the peak fallback rate and accretion luminosity are
\begin{align}
  \dot M_{\rm peak} &= \frac{M_p}{3t_{\rm peak}} = 6.58\times10^{-6}\,\Msun\,{\rm yr}^{-1}, \label{eq:Mdot_peak_blanet}\\
  L_{\rm peak} &= \eta\dot M_{\rm peak}c^2 = 3.7\times10^{40}\,{\rm erg\,s^{-1}} = 3.0\times10^{-5}\,\Ledd. \label{eq:L_peak}
\end{align}
The peak disc temperature, using the standard thin-disc scaling, is
\begin{equation}
  T_{\rm max}\approx2.3\times10^5\,{\rm K}
  \left(\frac{\MBH}{10^7\Msun}\right)^{-1/4}
  \left(\frac{\dot M_{\rm peak}}{\Medd}\right)^{1/4}
  \approx1.7\times10^4\,{\rm K},
  \label{eq:T_max}
\end{equation}
peaking near $\lambda\approx171$~nm in the far-UV. Applying the same formal thin-disc scaling to the Sun-like stellar event gives $T_{\rm max}\approx3.3\times10^5$~K. These estimates assume blackbody emission from circularised debris; the stellar value is additionally subject to the limitations of applying a thin-disc scaling to a super-Eddington peak fallback rate. Reprocessing by a pre-existing AGN flow and strong-field hydrodynamic effects, including stream self-intersection, shocks, and relativistic compression, may modify the luminosity and temperature normalisations. Modelling these processes requires relativistic hydrodynamic calculations such as those of \citep{Tejeda2017,GaftonRosswog2019}.

\subsection{Flare duration and survey strategy}

The total flare duration, $\Delta t_{\rm flare}\sim10\,t_{\rm peak}$, is therefore
\begin{equation}
  \Delta t_{\rm flare} \approx 1.5\times10^2\,{\rm yr},
  \label{eq:duration}
\end{equation}
This long duration has direct consequences for detectability. A century-scale event overlaps with long-term intrinsic AGN variability and extends well beyond the baseline of individual time-domain surveys. The $t^{-5/3}$ decay would appear as a decades-long secular trend in archival photometry rather than as a rapidly evolving transient. Detection therefore favours multi-survey archival analyses of AGN with independently constrained low-mass SMBHs. Host contamination remains a major challenge because the $\sim10^{40}\,{\rm erg\,s^{-1}}$ signal is superposed on stochastic AGN variability over overlapping timescales.

Table~\ref{tab:comparison} collects the principal physical differences between the Sun-like stellar event and the fiducial blanet event at the same SMBH mass.

\begin{table}[htbp]
\centering
\caption{Stellar vs.\ blanet TDE properties, self-consistently evaluated at $\MBH=10^7\,\Msun$.}
\label{tab:comparison}
\begin{tabular}{@{}lcc@{}}
\toprule
Quantity & Sun-like star & Fiducial blanet \\
\midrule
Body mass $M_p$                  & $1\,\Msun$ & $100\,\Me$ \\
Body radius $R_p$                & $1\,\Rsun$ & $6\,\Re$ \\
$\rt$ (AU)                      & $1.00$      & $0.822$ \\
$\rt/\rs$                       & $5.08$      & $4.16$ \\
$t_{\rm peak}$                  & $129.5$~d   & $15.2$~yr \\
$\dot M_{\rm peak}$ ($\Msun\,{\rm yr^{-1}}$) & $0.94$ & $6.58\times10^{-6}$ \\
$T_{\rm max}$ (K)              & $\sim3.3\times10^5$ & $1.7\times10^4$ \\
$M_{\rm Hills}$ ($\Msun$)       & $1.14\times10^8$ & $8.50\times10^7$ \\
\bottomrule
\end{tabular}
\end{table}

Raw sensitivity is not, by itself, the limiting factor. Using $F=L_{\rm peak}/4\pi D_L^2$ and approximating the UV spectrum's flux density at the Wien peak, Table~\ref{tab:detect} gives the resulting flux and rough AB magnitude at three representative distances against approximate single-visit/coadd depths for two relevant facilities. These are order-of-magnitude estimates, not a substitute for a proper synthetic-photometry calculation, but they show that duration and host contamination (discussed above), not depth, are the binding observational constraints.

\begin{table}[htbp]
\centering
\caption{Indicative comparison of the peak-luminosity flux with representative survey depths. The estimates neglect host contamination, extinction, bandpass corrections, and redshift effects.}
\label{tab:detect}
\begin{tabular}{@{}lcc>{\raggedright\arraybackslash}p{0.36\columnwidth}@{}}
\toprule
$D_L$ & \shortstack{$F_{\rm bol}$\\(erg\,s$^{-1}$\,cm$^{-2}$)} & \shortstack{Approx.\\AB mag} & Representative survey detectability \\
\midrule
10~Mpc  & $3\times10^{-12}$ & $\sim18$ & Brighter than representative Rubin and Roman depths \\
100~Mpc & $3\times10^{-14}$ & $\sim23$ & Within representative Rubin and Roman depths \\
500~Mpc & $1\times10^{-15}$ & $\sim26$ & Near the Rubin coadd depth and within the representative Roman depth \\
\bottomrule
\end{tabular}
\end{table}

\begin{figure}[htbp]
  \centering
  \includegraphics[width=\columnwidth]{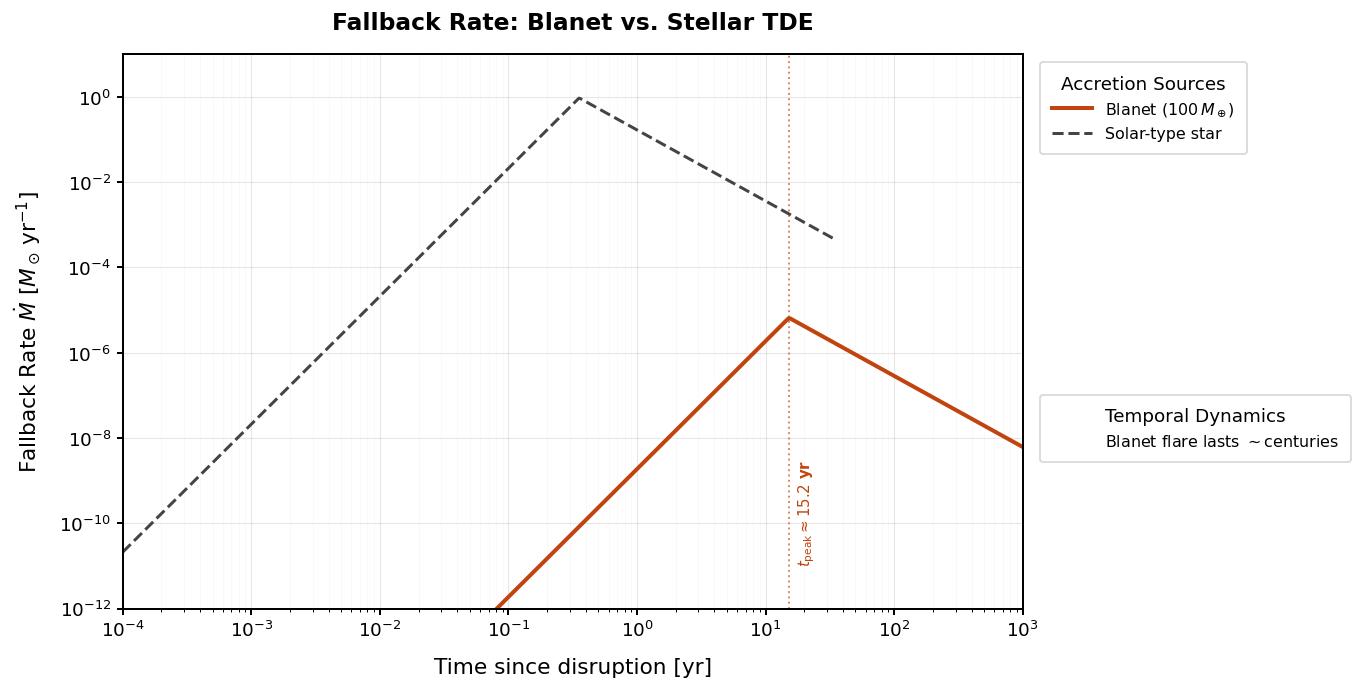}
  \caption{Schematic illustration of the fallback rate as a function of time since disruption for the fiducial blanet and a Sun-like star, both at $\MBH=10^7\,\Msun$. The straight pre-peak segments are visual guides rather than structure-dependent calculations of the rise to peak; each curve is normalized at its respective peak fallback time and follows the canonical $t^{-5/3}$ decline thereafter. The blanet fallback rate peaks at $t_{\rm peak}\approx15.2$~yr, substantially later and at a much lower rate than the stellar event, and its decline extends over century timescales.}
  \label{fig:fallback}
\end{figure}

Figure~\ref{fig:fallback} compares the fallback evolution of the fiducial blanet with that of a Sun-like star disrupted by the same SMBH.

\FloatBarrier

\section{Gravitational-wave prospects for post-disruption debris}\label{sec:gw}

A fraction of the disrupted blanet may remain bound in self-gravitating fragments that circularise and undergo extreme-mass-ratio-inspiral-like evolution. We estimate the gravitational-wave signal at $r_0\approx\rt\approx8.3\,M$ (Section~\ref{subsec:hills}). Because this radius is close to the ISCO, stable circular motion is not possible for every spin and inclination. Viscous spreading may move debris outward before an adiabatic inspiral develops, so $r_0=\rt$ represents a lower bound on the relevant orbital radius.

At $r_0=8.3\,M$, the GW frequency is
\begin{equation}
  f_{\rm GW} = 2f_{\rm orb} \approx 2.7\times10^{-4}\,{\rm Hz},
  \label{eq:fGW}
\end{equation}
within LISA's approximate $0.1$~mHz--$1$~Hz band. Using the standard sky-averaged circular-orbit strain amplitude,
\begin{equation}
  h_0 = \sqrt{\frac{32}{5}}\,\frac{G^2}{c^4}\,\frac{\MBH m_f}{D_L\,r_0},
  \label{eq:hc}
\end{equation}
For a fragment mass $m_f=0.01\,\Me$ at $D_L=100$~Mpc, Eq.~(\ref{eq:hc}) gives $h_0\approx4\times10^{-30}$, approximately nine orders of magnitude below plausible LISA sensitivity ($\sim10^{-21}$--$10^{-20}$ at comparable frequencies). The suppression is driven by the fragment mass rather than the signal frequency. The scaling $h_0\propto \MBH m_f D_L^{-1}r_0^{-1}$ shows that detectability would require an implausibly large increase in $m_f/D_L$. The same mass-dependent limitation applies to other planned mHz detectors, including Taiji and TianQin. Eccentric harmonics may redistribute the signal power but cannot overcome the amplitude deficit. As a consistency check, Eq.~(\ref{eq:hc}) gives $h_0\sim10^{-22}$ for a $10\,\Msun$ compact object at $r_0=10M$ around a $10^6\,\Msun$ SMBH at $D_L=1$~Gpc, consistent with the expected scale for a stellar-mass EMRI.

Comparable-mass tidal-phase expressions \citep{Flanagan2008} are not applicable to an extreme-mass-ratio debris fragment without a dedicated derivation of its tidal deformability. The robust result is therefore the mass threshold implied by Eq.~(\ref{eq:hc}): even fragments approaching the blanet's total mass would remain far below single-event detectability at extragalactic distances.

\FloatBarrier

\section{Event rate}\label{sec:rate}

Adopting scalar resonant relaxation (Section~\ref{subsec:destabilisation}) as the principal delivery channel, with $t_{\rm SRR}\sim10^8$--$10^9$~yr and an AGN episode lifetime $\tau_{\rm AGN}\sim10^8$~yr, gives a destabilised fraction $f_{\rm SRR}\sim\tau_{\rm AGN}/t_{\rm SRR}\sim0.01$--$0.1$ for a population of $N_{\rm blanet}\sim10^3$. The resulting per-AGN rate is
\begin{equation}
  \dot N_{\rm TDE} = \frac{N_{\rm blanet}f_{\rm SRR}}{\tau_{\rm AGN}} \sim 10^{-7}\text{--}10^{-6}\,{\rm yr^{-1}\,AGN^{-1}},
  \label{eq:blanet_rate}
\end{equation}
Using an AGN space density $n_{\rm AGN}\sim10^{-4}$--$10^{-3}\,{\rm Mpc^{-3}}$, the corresponding volumetric rate is
\begin{equation}
  \mathcal{R}_V \sim 10^{-11}\text{--}10^{-9}\,{\rm Mpc^{-3}\,yr^{-1}},
  \label{eq:vol_rate}
\end{equation}
approximately $10^4$--$10^7$ times below the stellar TDE rate of $\sim10^{-5}$--$10^{-4}\,{\rm Mpc^{-3}\,yr^{-1}}$.

\FloatBarrier

\section{Discussion and conclusions}\label{sec:discussion}

The results define a physical picture of blanet TDEs with three closely connected features.

First, disruption occurs in the strong-field regime. For the fiducial parameters, $\rt\approx8.3\,\rg$, comparable to the ISCO, and the Kerr effects described in Section~\ref{sec:tidal} must be treated non-perturbatively.

Second, the Hills mass is a density diagnostic rather than a uniform enhancement relative to stellar TDEs. Rocky blanets can be disrupted over a smaller SMBH-mass range than Sun-like stars, whereas volatile-rich blanets can be disrupted over a larger range. Blanet TDE demographics may therefore constrain bulk composition.

Third, the observable signature is a faint ($\sim3\times10^{-5}\,\Ledd$), UV-peaked transient lasting $\sim10^2$~yr. Together with the low event rate associated with scalar resonant relaxation, this favours archival, multi-decade searches and indicates that blanet TDEs may constitute a rare but potentially distinctive transient class.

Although the present treatment is analytical, existing relativistic hydrodynamic simulations provide useful context for its interpretation. The stellar-disruption calculations of \citep{Tejeda2017} show that black-hole spin can strongly reshape the debris streams while having comparatively little effect on the fallback rate in the cases they considered, whereas \citep{GaftonRosswog2019} find that deep or partial relativistic encounters can also modify the fallback normalisation, rise, peak time, and return times. These simulations concern stellar rather than planetary disruptions and therefore do not directly validate the fiducial blanet calculation; instead, they support interpreting the analytical results as zeroth-order dynamical scales. Detailed debris morphology, circularisation, and emission remain questions for dedicated relativistic hydrodynamic simulations.

Individual post-disruption debris fragments are not promising gravitational-wave sources. Even at the near-ISCO disruption radius, the strain from sub-Earth-mass fragments is far below LISA sensitivity because of their small mass.

\subsection{Validity domain and limitations}

The analytical framework applies in the small-mass-ratio regime ($q\ll1$) and adopts ballistic, non-self-gravitating debris in the frozen-in approximation. It assumes a vacuum Kerr spacetime and neglects debris back-reaction, self-force, and MPD spin-curvature coupling, the latter being small for $q\sim10^{-8}$ as discussed in Section~\ref{sec:geodesics}. The emission model assumes blackbody radiation without reprocessing by a pre-existing AGN flow. Radiation-hydrodynamic coupling, stream self-intersection, shocks, and relativistic compression during circularisation are also omitted. These effects may alter the predicted luminosity and temperature, particularly because disruption occurs near the ISCO, and require relativistic hydrodynamic calculations \citep{Tejeda2017,GaftonRosswog2019}. The efficiency of scalar resonant relaxation additionally depends on the nuclear star cluster density profile and coherence time.

The Type-I migration and disc-precession timescales in Section~\ref{subsec:destabilisation} depend on the AGN disc surface-density profile and should be evaluated within a specified disc model, such as those implemented in \texttt{pAGN} \citep{Gangardt2024}. Population-level predictions also require the prograde and retrograde distributions of the orbital constants $L$ and $Q$, rather than the representative $L/M$ range used here. Table~\ref{tab:summary} collects the principal fiducial predictions.

\begin{table}[htbp]
\centering
\caption{Fiducial blanet TDE predictions ($M_p=100\,\Me$, $R_p=6\,\Re$, $\MBH=10^7\,\Msun$), collected for reference.}
\label{tab:summary}
\begin{tabular}{@{}lc@{}}
\toprule
Quantity & Value \\
\midrule
Tidal radius $\rt$ & $0.82$~AU ($8.3\,\rg$) \\
Hills mass $M_{\rm Hills}$ (this composition) & $8.5\times10^7\,\Msun$ \\
Peak fallback time $t_{\rm peak}$ & $15.2$~yr \\
Peak luminosity $L_{\rm peak}$ & $3.7\times10^{40}\,{\rm erg\,s^{-1}}$ ($3\times10^{-5}\Ledd$) \\
Peak temperature $T_{\rm max}$ & $1.7\times10^4$~K ($\sim171$~nm) \\
Flare duration $\Delta t_{\rm flare}$ & $\sim150$~yr \\
GW strain $h_0$ ($m_f=0.01\,\Me$, $r_0\approx\rt$, $D_L=100$~Mpc) & $\sim4\times10^{-30}$ \\
Per-AGN rate $\dot N_{\rm TDE}$ & $10^{-7}$--$10^{-6}\,{\rm yr^{-1}}$ \\
\bottomrule
\end{tabular}
\end{table}

\FloatBarrier

\noindent\textbf{Summary of fiducial predictions.} Taken together, the benchmarks in Table~\ref{tab:summary} describe disruption near the ISCO followed by a delayed, century-scale, low-luminosity UV transient with no viable single-event gravitational-wave counterpart. These results may provide concrete quantitative targets for future time-domain searches and relativistic hydrodynamic simulations of planetary tidal disruption events in AGN environments.

\section*{Acknowledgements}
The authors gratefully acknowledge the Department of Physics, Kirori Mal College, University of Delhi, and thank the Dr.~N.~S.~Pradhan Memorial Library and the CARPA Library, Department of Physics and Astrophysics, University of Delhi, for access to their academic resources. We also sincerely thank the anonymous referee and the editor for their constructive comments and suggestions, which substantially improved the clarity and quality of the manuscript.

\appendix
\section{Tidal-tensor construction and limiting cases}\label{app:derivation}

This appendix summarises the construction of the tidal eigenvalues used in Section~\ref{subsec:kerr_tidal} and their limiting behaviour, following \citep{Marck1983} and its application to Kerr disruption calculations by \citep{Kesden2012}. For equatorial geodesics, the explicit intermediate algebra of projecting the Riemann tensor onto the Marck tetrad and diagonalizing the resulting tidal matrix follows the specialization used by \citep{Kesden2012} and is not reproduced in full; the essential geometric construction and limiting checks are given below.

\textbf{Setup.} For a geodesic with four-velocity $u^\mu$, nearby geodesics separated by $\xi^\mu$ obey $D^2\xi^\mu/d\tau^2=-R^\mu{}_{\nu\alpha\beta}u^\nu u^\alpha\xi^\beta$. In an orthonormal tetrad $\{e_{\hat 0}=u,e_{\hat 1},e_{\hat 2},e_{\hat 3}\}$ comoving with the geodesic, the spatial tidal tensor is $C_{\hat\imath\hat\jmath}=R_{\hat\imath\hat 0\hat\jmath\hat 0}$, and disruption occurs along the eigenvector of the most negative eigenvalue (radial stretching).

\textbf{Marck tetrad.} A tetrad that is both orthonormal and parallel-transported ($De_{\hat a}^\mu/d\tau=0$) is required so that $C_{\hat\imath\hat\jmath}$ is evaluated in a genuinely non-rotating comoving frame along the true (generally eccentric) orbit, rather than a locally-non-rotating (ZAMO) frame that is not comoving with the particle. Marck's construction uses the Kerr spacetime's Killing-Yano tensor to build such a tetrad explicitly in terms of $E$, $L$, $Q$, and $r(\tau)$, $\theta(\tau)$ along the geodesic, without needing to integrate a separate parallel-transport equation.

\textbf{Reduction for equatorial orbits.} For $\theta=\pi/2$ throughout (equatorial motion, $Q$ enters only through $K$), the tetrad simplifies and the Riemann tensor components project onto it to give the diagonal result quoted in Eqs.~(\ref{eq:Lambda1})--(\ref{eq:Lambda3}), with the combination $K=(L-a_*E)^2+Q$ appearing because it is precisely the combination that enters the radial potential $\mathcal{R}(r)$ (Eq.~\ref{eq:calR}) and therefore the second radial derivative of the effective potential that the tidal tensor measures.

\textbf{Schwarzschild limit.} Setting $a=0$ gives $K\to L^2$. For radial infall ($L=0$), the eigenvalues become $\Lambda_1=2M/r^3$ and $\Lambda_2=\Lambda_3=-M/r^3$ at every $r$, reproducing Eq.~(\ref{eq:tidal_schw}). An independent ZAMO-frame derivation for stationary axisymmetric spacetimes yields the same Schwarzschild and Kerr limits \citep{XinMummery2026}.

\begin{sloppypar}

\end{sloppypar}

\end{document}